\documentclass[sigconf,nonacm]{acmart}
\usepackage{booktabs}
\usepackage{tabularx}
\usepackage{newunicodechar}
\usepackage{makecell}
\newunicodechar{→}{$\rightarrow$}
\newunicodechar{×}{$\times$}
\newunicodechar{≈}{$\approx$}
\newunicodechar{≥}{$\geq$}
\newunicodechar{≤}{$\leq$}
\newunicodechar{−}{$-$}
\newunicodechar{–}{--}
\newunicodechar{—}{---}
\newunicodechar{§}{\S}
\newunicodechar{ρ}{$\rho$}
\newunicodechar{α}{$\alpha$}
\newunicodechar{λ}{$\lambda$}
\newunicodechar{μ}{$\mu$}
\newunicodechar{κ}{$\kappa$}
\newunicodechar{≠}{$\neq$}
\newunicodechar{⇒}{$\Rightarrow$}
\newunicodechar{“}{``}
\newunicodechar{”}{''}
\newunicodechar{‘}{`}
\newunicodechar{’}{'}
\newunicodechar{é}{\'e}
\newunicodechar{紧}{jin}
\newunicodechar{急}{ji}
\newunicodechar{严}{yan}
\newunicodechar{重}{zhong}
\newunicodechar{要}{yao}
\newunicodechar{•}{\textbullet}
\providecommand{\tightlist}{\setlength{\itemsep}{0pt}\setlength{\parskip}{0pt}}

\begin{document}
\raggedbottom

\title{Rethinking Vulnerability Remediation as a Capacity Allocation Problem}
\author{J. Stucke}
\affiliation{%
  \city{Munich}
  \country{Germany}
}

\begin{abstract}
As AI accelerates vulnerability discovery, remediation throughput may become a greater constraint than prioritisation accuracy. This study evaluates vulnerability remediation as a flow-control problem using Apache Jira, Mozilla Bugzilla, Red Hat security errata, five public Jira organisations, and an npm dependency graph.

Apache resolution times are strongly heavy-tailed (Hill $\alpha=2.24$; mean-to-median ratio $10.7\times$), while 94--100\% of arrivals in the primary issue trackers enter queues estimated at $\rho \geq 1$. Queue-context models achieve only moderate discrimination (AUC 0.66--0.69), largely matched by simple project-level baselines, and no evaluated conformal deadline method outperforms a stratified empirical quantile. Severity-to-speed discrimination increases from Apache (AUC 0.505) to Mozilla (0.643) and Red Hat (0.785), although this cross-system gradient is observational.

Flow-control interventions show larger operational effects. Two Apache transitions from overloaded to draining queues are associated with 90--116 fewer resolution days across two control specifications. At fixed capacity, severity-first sequencing reduces critical-item dwell by 227 days in Apache and 59 days in Mozilla. For Apache, reserving 25.4\% of capacity for critical work yields a model-estimated mean sojourn of 137.5 days compared with an observed mean of 265 days. Owner-level analysis further shows that capacity cannot always be transferred where needed: expertise-eligible borrowing changes critical-item feasibility by 0.0--85.3 percentage points across five Jira organisations, revealing siloed, connected but saturated, connected with slack, and already provisioned regimes.

Finally, npm vulnerabilities cluster weakly along dependency relationships ($+0.083$, $p=0.011$), indicating that exposure is not fully item-independent. Together, these observational and model-based findings show that remediation performance depends on queue state, capacity allocation, ownership, and expertise connectivity, and support treating vulnerability remediation as a flow-control rather than solely a ranking problem.

\end{abstract}

\keywords{vulnerability remediation, mining software repositories, queueing
theory, flow control, issue trackers, prioritisation, conformal
prediction, capacity reservation.}

\maketitle

\section{Introduction}\label{introduction}

Vulnerability-management systems typically operationalise urgency through severity scores, exploitation probabilities, and mandated deadlines. CVSS summarises technical severity, EPSS estimates exploitation probability~\cite{ref4}, and CISA BOD 26-04 prescribes risk-dependent remediation timelines~\cite{ref17}. These mechanisms indicate which vulnerabilities should be addressed first, but they do not determine whether the responsible owner has sufficient capacity, whether authorised and capable assistance is available, or whether aggregate arrival rates exceed remediation throughput.
\begin{sloppypar}
This study formulates vulnerability remediation as a flow-control problem. The empirical analysis integrates queueing diagnostics, predictive baselines, cross-organisational comparisons, owner-\discretionary{}{}{}constrained capacity models, resolver-overlap networks, observed queue-state transitions, counterfactual sequencing analyses, and dependency-graph structure. The central claim is intentionally limited in scope. Within the systems examined, operational delay is more directly associated with queue conditions and deployable, owner-constrained capacity than with additional sophistication in prioritisation scores. The analysis makes five contributions:
\end{sloppypar}
\begin{itemize}
\tightlist
\item
  \textbf{C1 --- Queue diagnosis.} Apache resolution times are
  heavy-tailed (Hill α = 2.24; mean/median 10.7×). An estimated
  94--100\% of arrivals in the primary tracker samples enter queues with
  ρ ≥ 1, where aggregate finite-delay steady state does not exist under
  the stationary model. Observed Little's-law quantities are strongly
  aligned across project-level time windows (r = 0.73--0.83), and both OSS trackers
  exhibit an approximately 1.8× dead-WIP factor.
\item
  \textbf{C2 --- Limited incremental value from prediction.} Apache
  priority discriminates slow from fast resolutions at AUC 0.505 (95\%
  CI 0.474--0.535). Queue-context features reach AUC 0.66--0.69, but a
  project-median lookup captures most of that signal. Conformal
  deadlines achieve out-of-time coverage, yet no evaluated conformal
  method beats a stratified empirical quantile under the one-sided
  Winkler score.
\item
  \textbf{C3 --- A governance-associated gradient.} Using a common temporal evaluation protocol, the severity-to-speed AUC increased across the three datasets, from 0.505 for Apache to 0.643 for Mozilla and 0.785 for Red Hat. This observational pattern is consistent with severity labels having greater operational influence in settings with stronger process enforcement, but it does not show that governance differences caused the increase.
\item
  \textbf{C4 --- Four regimes of capacity mobility.} Using a consistent definition of ownership across five public Jira organisations~\cite{ref18}, the analysis shows that organisational category alone does not indicate whether cross-owner borrowing will be beneficial. Instead, the combination of resolver connectivity and available helper capacity distinguishes four operating conditions: siloed, connected but saturated, connected with slack, and already provisioned. With borrowing capped at 25\%, the increase in critical-item feasibility ranged from 0.0 to 85.3 percentage points.
\item
  \textbf{C5 --- Measurable flow levers.} Two observed queue-state transitions were associated with resolution times that were 90--116 days shorter across two alternative control constructions. In a counterfactual replay that kept total capacity unchanged, processing higher-severity items first reduced the estimated time that critical items remained in the queue by 227 days for Apache and 59 days for Mozilla. For the Apache dataset, a separate fast-track model estimated that reserving 25.4\% of effective capacity for critical items yields a mean time in the system of 137.5 days, compared with the observed mean of 265 days. Since these findings are based on observational comparisons and modelled counterfactuals, they should not be interpreted as randomised causal effects.
\end{itemize}

A supporting npm analysis finds positive vulnerability
assortativity (+0.083, p = 0.011) and base-before-dependent temporal
ordering in 60\% of eligible pairs (p = 0.019). These results are
consistent with network-structured exposure but do not by themselves
demonstrate propagation.

These findings do not suggest that severity or exploitation scores are unnecessary. EPSS, for example, reports a ROC AUC of 0.838 for predicting exploitation \cite{ref4}. Scoring helps determine which items should be addressed first, whereas throughput, ownership, expertise, and change constraints determine whether that priority order can be carried out. The study also does not claim causal field evidence for every intervention. Each result is explicitly identified as descriptive, quasi-experimental, backtested, or analytical.

\section{Related Work}\label{related-work}
Research on vulnerability remediation has largely followed three complementary lines. Scoring approaches assess risk and support prioritisation, predictive methods estimate remediation time, and queueing and workflow models examine how capacity constraints shape completion. This section reviews these strands of work and situates the present study within the remaining gap.

\paragraph{Risk scoring and remediation prioritisation.}
The Common Vulnerability Scoring System (CVSS) ranks vulnerabilities by technical severity, whereas the Exploit Prediction Scoring System (EPSS) estimates the likelihood that a vulnerability will be exploited \cite{ref4}. The US Cybersecurity and Infrastructure Security Agency (CISA) further translates risk assessments into remediation requirements through Binding Operational Directive 26-04 \cite{ref17}. Together, these approaches help organisations identify urgent vulnerabilities and define when they should be addressed. They do not, however, account for whether the responsible teams have sufficient capacity, expertise, or operational flexibility to complete the work within those timelines. In this study, risk scores are therefore treated as prioritisation signals rather than as complete remediation strategies.

\paragraph{Fix-time prediction.}
A separate line of research estimates how long software issues will take to resolve. Previous studies have used similarity between issues \cite{ref1}, attributes extracted from bug reports \cite{ref2}, and broader process information to predict resolution effort or duration. Ruohonen \cite{ref3} found that reporter identity explained variation in CPython handling times, whereas vulnerability severity and the presence of proof-of-concept code did not. These findings suggest that organisational and workflow context may be more informative than risk labels alone. This study therefore examines whether such predictive approaches generalise across different governance settings and outperform simple queue-aware baselines.

\paragraph{Queueing and flow control.}
Previous studies have modelled vulnerability backlogs and attack surfaces as queues. Heavy-tailed patching times can produce persistent backlogs \cite{ref5}, vulnerability timestamps can support estimates of required resources \cite{ref6}, and faster vulnerability discovery can exceed remediation capacity in simulated interconnected systems \cite{ref7}. The present study builds on this work by using issue-tracker data to characterise queue behaviour and evaluate specific flow-control mechanisms.
Related flow-control mechanisms have also been studied in other operational settings. Emergency-department fast tracks reserve capacity for selected patient groups and have shown moderate evidence of reducing waiting times \cite{ref10}. In software engineering, a Kanban case study linked lower work in progress to shorter lead times, while showing that productivity effects depend on context \cite{ref11}. Classical priority-queue theory provides formulas, including Cobham's, for estimating waiting times when higher-priority work is served first without interrupting work already in progress \cite{ref9}. Akbarzadeh and Mahajan (2023) further showed that partially observable restless-bandit models can support resource allocation when item-level state transitions can be estimated reliably \cite{ref12}. Building on these approaches, the present study applies capacity reservation to vulnerability queues and evaluates its feasibility analytically rather than through a deployed intervention.

\paragraph{Ownership, expertise, and reassignment.}
Jeong et al.~found that 37\%--44\% of Mozilla and Eclipse bug reports were reassigned, with reassignment generally associated with longer resolution times \cite{ref13}. Guo et al.~showed, however, that reassignment can also help identify the correct owner, relevant expertise, root cause, or available capacity \cite{ref14}. Together, these findings suggest that transferring work can be useful, but may introduce coordination costs. The present study therefore models both resolver expertise and the overhead associated with reassignment.

\paragraph{Uncertainty-aware deadlines.}
Split conformal prediction \cite{ref15} and conformalised quantile regression (CQR) \cite{ref16} provide prediction ranges with statistical coverage guarantees without requiring a specific probability distribution. In this study, these methods are compared with simpler empirical and parametric baselines to assess whether improved statistical calibration also translates into more useful remediation deadlines in practice.

\paragraph{Public Jira data.}
Montgomery et al.~released a catalogue and dataset covering 16 public Jira instances, 1,822 projects, and 2.7 million issues \cite{ref18}. Five organisations from this catalogue are used to evaluate capacity mobility at a common ownership granularity. Fresh public REST records are retrieved and cached to support deterministic reruns.

\section{Data and Methodology}\label{data-and-methods}

To test whether remediation performance is constrained more by prioritisation or by the underlying flow of work, the study combines public issue-tracking records, vulnerability disclosures, and software-dependency data. The analysis examines remediation at three complementary levels. Queue-level methods characterise workload, utilisation, and delay; owner-level methods assess whether remediation capacity can be redistributed across teams subject to expertise constraints; and dependency-level methods examine whether remediation patterns are structured by relationships between software packages. All datasets are public and are stored in fixed caches to enable deterministic reproduction of the analyses.

\paragraph{Primary corpora.}
The queue, prediction, and governance analyses draw on four public datasets. These comprise 2,000 Apache Jira issues from multiple projects, of which 1,672 are resolved; 2,000 Mozilla Bugzilla Core bugs; 163 Red Hat Common Vulnerabilities and Exposures (CVE) records measuring the time from public disclosure to the first Red Hat Security Advisory (RHSA) and labelled by threat severity; and an npm dependency graph containing 1,021 packages with Open Source Vulnerabilities (OSV) advisory histories.

\paragraph{Cross-organisation Jira sample.}
Capacity mobility is analysed across five public Jira organisations—Apache, MariaDB, Qt, MongoDB, and Red Hat—using Jira projects as a common proxy for ownership. Each organisation contributes 2,000 issues. Apache uses the longer-span primary cached sample, whereas the remaining four datasets contain issues created on or after 1 January 2022 and were retrieved through anonymous public REST endpoints. The organisations were selected from the public Jira catalogue compiled by Montgomery et al.~\cite{ref18}.

Because priority schemes differ across Jira instances, labels are normalised for each organisation. The anonymous Red Hat endpoint returns localised priority labels, which are mapped to blocker, critical, and major. Differences in priority schemes, observation windows, and the use of equal sample sizes across organisations are considered potential threats to validity.

\paragraph{Queue construction.}
Jira projects and Mozilla components are used as proxies for remediation queues and their responsible owners. Queue utilisation at the time each ticket is created, $\rho$, is estimated from arrival and completion rates. A queue is classified as \textbf{supercritical} when $\rho \geq 1$, meaning that work arrives at least as fast as it is completed, and as \textbf{draining} when $\rho < 1$. Under a stationary aggregate queue model, $\rho \geq 1$ indicates that no stable steady state with finite average delay exists. However, it does not imply that every individual priority class must experience infinite delay.

\paragraph{Diagnostics and prediction.}
Queue behaviour and remediation times are characterised using several complementary measures. The Hill estimator \cite{ref8} is used to assess the heavy-tailed behaviour of resolution times and is validated against synthetic Pareto data with known properties. Consistency with Little's law is evaluated across project-level time windows. The analysis tests whether vulnerability severity can distinguish issues that are resolved quickly from those that take longer. Performance is measured using the area under the receiver operating characteristic curve (AUC) on later, unseen data, with bootstrap confidence intervals used to quantify uncertainty.

Remediation-deadline methods are evaluated using a temporal split in which 40\% of the data are used for training, 30\% for calibration, and the final 30\% for testing on later cases. The comparison includes simple empirical-quantile baselines, a log-normal accelerated failure-time model, and several conformal approaches, including conformal prediction over ordinary least squares, conformalised quantile regression (CQR), and a Mondrian method that accounts for queue state.
Coverage is reported with Wilson 95\% confidence intervals, while the usefulness of the predicted deadlines is evaluated using a one-sided Winkler score that accounts for both missed deadlines and unnecessarily conservative predictions.

\paragraph{Owner and expertise constraints.}
Owner-level remediation feasibility is estimated from the arrival rate of critical items and the observed processing throughput of each owner. Resolver identities are used as a proxy for demonstrated expertise. One owner is considered eligible to support another when their resolver sets share at least two identities or have a Jaccard similarity of at least 0.1.

Eligible owners may contribute only residual capacity available under the same reservation limit, and each unit of spare capacity can be allocated once. Missing assignee information introduces uncertainty because it may underestimate overlap between resolver groups and may be systematically distributed rather than random.

\paragraph{Evaluation of flow-control mechanisms.}
Historical transitions from supercritical to draining queue states are evaluated using difference-in-differences comparisons with contemporaneous queues that did not undergo the same transition. A separate sequencing replay holds measured capacity constant and evaluates how alternative ordering of work affects critical-item delay within the same owner.

The fast-track model reserves a fraction $f$ of effective capacity for critical items. Expected sojourn time in the reserved lane is estimated using Pollaczek--Khinchine quantities and Cobham's priority-queue framework \cite{ref9}. All model parameters are derived from observed tracker data. Because no fast-track intervention was deployed, the resulting estimates are analytical counterfactuals rather than observed causal effects.

\section{Results}
\subsection{Queue Diagnosis}\label{queue-diagnosis}

This study first examines whether remediation delay is consistent with capacity and queueing constraints rather than with prioritisation alone. The results show long-tailed resolution times, persistently high queue utilisation, and systematic relationships between backlog and remediation delay, providing empirical support for a flow-based interpretation of remediation performance.

\textbf{Heavy-tailed resolution time.} Among the 1,672 resolved Apache issues, resolution times are strongly heavy-tailed. The Hill tail-index estimate is $\alpha = 2.24$ \cite{ref8}, and the mean resolution time is 10.7 times larger than the median. Because the estimate lies close to the $\alpha = 2$ boundary at which the variance becomes highly sensitive to extreme observations, rare long-running issues can strongly influence second-moment statistics. The analysis therefore reports medians where possible and uses winsorised sensitivity checks for calculations that depend on second moments.

\textbf{High estimated utilisation.} Approximately 94\% of Apache arrivals and nearly all Mozilla Core arrivals enter queues with estimated utilisation $\rho \geq 1$. In these queues, new work arrives at least as fast as existing work is completed, and observed resolution times are approximately three times longer than in draining queues. Under the stationary aggregate queue model, such supercritical queues cannot sustain a stable finite-delay steady state. Reordering work may still protect a sufficiently small high-priority class, but it cannot remove the aggregate backlog or increase total remediation throughput.

\textbf{Observable flow regularity.} Across project-level time windows, observed backlog $L$ is strongly correlated with the quantity $\lambda W$ ($r = 0.73$--$0.83$), where $\lambda$ denotes the arrival rate and $W$ the average time spent in the system. This pattern is consistent with Little's law, although the observed correlation does not by itself establish all assumptions required by the underlying stationary model. Both open-source trackers also show an approximately $1.8\times$ dead-work-in-progress factor for items that remain open and inactive for at least one year. Together, these patterns support the use of queue-based diagnostics for analysing remediation delays, although they do not establish that queue dynamics alone explain the observed differences in resolution time.

\subsection{Limited Incremental Value of
Prediction}\label{limited-incremental-value-of-prediction}

The queue diagnostics show that remediation time is strongly associated with project and queue conditions. The analysis therefore examines whether predictive modelling provides additional value beyond this baseline.

\textbf{Priority provides negligible discrimination on Apache.}
Apache priority labels provide almost no ability to distinguish issues that are resolved quickly from those that take longer. Using a temporal train--test split, the resulting AUC is 0.505 (95\% CI 0.474--0.535), with the confidence interval including the 0.5 level expected from random discrimination. Under an earlier exploratory evaluation, an estimate based on only
500 issues produced an AUC of 0.613; this declined to 0.523 when that
analysis was expanded to 2,000 issues. The earlier result is therefore retained in the validation record as an illustration of small-sample instability rather than as evidence of predictive performance.

\textbf{Simple queue context captures most observed signal.} Models that incorporate project history and backlog at the time an issue arrives achieve AUC values between 0.66 and 0.69. However, a simple lookup based on the historical median resolution time of each project performs nearly as well. This suggests that most of the measurable predictive signal is associated with persistent differences between projects and queues, while additional model complexity provides only limited improvement.

\textbf{Coverage does not imply efficient deadlines.} Split-conformal methods \cite{ref15} achieve out-of-time coverage close to or above the nominal 90\% target. Coverage reaches 93.4\% for Apache (Wilson 95\% CI 90.9--95.3\%, $n = 503$) and 93.7\% for Mozilla (91.3--95.5\%, $n = 510$). Across rolling-origin evaluations, mean coverage is 92.2\%, with a minimum of 89.6\%.

High coverage, however, does not necessarily imply useful remediation deadlines. Under the one-sided Winkler score, which also penalises excessively wide bounds, a non-conformal baseline performs best in every evaluated corpus-by-target comparison, while conformalised quantile regression (CQR) \cite{ref16} consistently outperforms the proposed Mondrian approach. In one rolling-origin evaluation, maintaining coverage requires an upper bound of 5,223 days. This illustrates that statistical coverage can be preserved by widening the prediction interval to a point at which the resulting deadline has little operational value.

These results highlight a limitation of uncertainty-aware prediction. Prediction bounds can widen sufficiently to preserve statistical coverage while becoming too broad to provide useful remediation deadlines. Such widening may nevertheless signal a change in the underlying remediation process or queue conditions. Bound width should therefore be considered both when assessing the operational usefulness of a prediction and as a potential indicator of distributional change.

\subsection{Governance-Associated Severity
Gradient}\label{governance-associated-severity-gradient}

The weak relationship between severity and resolution speed in Apache is not observed equally across all systems. Applying the same temporal evaluation protocol to three remediation settings shows that this relationship becomes stronger as the processes become more formally governed (Table~\ref{tab:severity-speed}).

\begin{table*}[!t]
\centering
\caption{Severity-to-speed discrimination measured by AUC with bootstrap 95\% confidence intervals.}
\label{tab:severity-speed}

\begin{tabularx}{\textwidth}{@{}lXXl@{}}
\toprule
System & Process setting & Signal & AUC (95\% CI) \\
\midrule
Apache Jira
& Volunteer project queues
& Tracker priority
& 0.505 (0.474--0.535) \\

Mozilla Bugzilla
& Corporate product triage
& Tracker severity
& 0.643 (0.606--0.678) \\

Red Hat errata
& SLA-oriented errata pipeline
& Threat severity
& 0.785 (0.702--0.859) \\
\bottomrule
\end{tabularx}
\end{table*}

Table~\ref{tab:severity-speed} shows that severity provides essentially
no discrimination in Apache, moderate discrimination in Mozilla, and
substantially stronger discrimination in the Red Hat errata process. The ordered pattern is consistent with the interpretation that severity becomes more closely associated with remediation speed when workflows more strongly enforce prioritisation.

However, this comparison remains observational and should not be interpreted as evidence that governance itself causes the increase in discrimination. The three systems differ in several other respects, including product context, workforce structure, ownership granularity, and data source, and the confidence intervals of adjacent systems overlap. Ruohonen's finding that severity was not associated with CPython handling time \cite{ref3} is consistent with the broader pattern that severity may be less consequential in comparatively decentralised remediation processes.

\subsection{Flow Levers and Capacity-Mobility
Regimes}\label{flow-levers-and-capacity-mobility-regimes}

The governance gradient suggests that severity labels are more closely associated with remediation speed when processes enforce their prioritisation. Prioritization alone, however, can only redirect existing remediation effort; it does not create additional capacity. The analysis therefore evaluates whether changes in sequencing and capacity allocation can reduce critical-item delay, and how ownership and expertise constrain their effectiveness.

\textbf{Draining transitions.}
Two Apache Jira project queues show sustained historical transitions from a supercritical state ($\rho \geq 1$) to a draining state ($\rho < 1$): MESOS in 2017 and FALCON in 2013. Median critical-item dwell decreased from 47 to 15 days in MESOS and from 277 to 71 days in FALCON, corresponding to a median treated change of $-119$ days. Relative to 13 non-flipping Apache project queues, the estimated reduction is 116 days when controls are evaluated over the same calendar windows and 90 days in a monthly-snapshot matched control replay. The 90--116-day range therefore reflects sensitivity to the construction of the control counterfactual rather than a confidence interval. Because only two treated transitions were observed, these estimates remain descriptive and should not be interpreted as a general causal effect.

\textbf{Fixed-capacity sequencing replay.} A counterfactual replay tests whether changing the order of work can improve critical-item resolution without increasing total capacity. Under severity-first sequencing, estimated critical-item dwell time decreases by 227 days in Apache and 59 days in Mozilla, assuming that remediation effort can be exchanged between items belonging to the same owner. Sequencing can therefore redistribute existing capacity towards critical work, but it does not increase overall throughput. Without an explicit allocation of capacity to critical items, ordering alone cannot guarantee that sufficient remediation effort remains available.

\textbf{Fast-track capacity reservation.} A stronger intervention is to reserve part of the available capacity specifically for critical work. Similar fast-track mechanisms are used in emergency departments, where systematic reviews provide moderate evidence of reduced waiting times \cite{ref10}. Here, a fraction $f$ of measured effective capacity is reserved for critical items, with effective capacity defined as $m_{\mathrm{eff}} = \mathrm{throughput} \times \mathrm{E}[S]$. Utilisation of the reserved lane is calculated as
$\rho_{\mathrm{lane}} = \lambda_c \mathrm{E}[S]/(f m_{\mathrm{eff}})$,
and mean sojourn time is estimated using Pollaczek--Khinchine quantities and Cobham's priority-queue framework \cite{ref9}. Critical arrival rates, throughput, and service-time moments are estimated from the observed tracker data. Resolution times from draining queues are used as a proxy for uncongested service time, with the second moment winsorised at the 99th percentile.
Using the Apache Jira data, Table~\ref{tab:apache-fast-track} shows how the estimated critical-item delay changes as a larger share of remediation capacity is reserved for critical work. With a 19.7\% reservation, the critical lane remains close to saturation at $\rho = 0.9$, producing an estimated mean sojourn of 640 days. Increasing the reservation to 25.4\% lowers utilisation to 0.7 and the estimated mean to 137.5 days, compared with the observed critical-item mean of 265 days. At a 35.5\% reservation, utilisation falls to 0.5 and the estimated mean sojourn to 46.7 days.

The Apache results therefore suggest that reserving capacity reduces critical-item delay only when a sufficiently large share is allocated to critical work. As utilisation of the reserved lane approaches full capacity, modelled waiting times increase non-linearly, making the choice of reservation level consequential. At the same time, capacity assigned to critical items is no longer available for non-critical work and may therefore increase delays elsewhere. Remediation capacity allocation therefore becomes an optimisation problem in which capacity must be distributed across competing classes of work to minimise critical delay without creating unacceptable congestion elsewhere.

\begin{table*}[!t]
\centering
\caption{Apache fast-track sensitivity (span 4,834 d; critical share 14.8\%;
$\mathrm{E}[S] = 123.3$ d; $m_{\mathrm{eff}} = 42.7$; measured critical
dwell median 17 d, mean 265 d).}
\label{tab:apache-fast-track}

\begin{tabularx}{\textwidth}{@{}llX@{}}
\toprule
Reservation $f$ & Lane $\rho$ & Model-estimated mean critical sojourn \\
\midrule
19.7\%          & 0.9          & 640 d \\
\textbf{25.4\%} & \textbf{0.7} & \textbf{137.5 d} \\
35.5\%          & 0.5          & 46.7 d \\
\bottomrule
\end{tabularx}
\end{table*}

The modelled benefit of capacity reservation primarily concerns the long-duration tail of critical-item remediation rather than the typical critical case. In Apache, critical items have a median dwell time of 17 days but a mean of 265 days, indicating a strongly right-skewed distribution in which a comparatively small number of long-running items substantially increase the average. The fast-track analysis therefore evaluates whether reserved capacity can reduce these prolonged delays rather than materially alter the median remediation time.
Because the capacity-reservation policy was not deployed operationally, the reported values represent analytical counterfactual estimates rather than observed treatment effects. The estimates are derived from a pooled M/G/1 queue model assuming Poisson arrivals, an empirically estimated service-time distribution, and non-preemptive processing, under which an item already in service is completed before a newly arriving critical item can be processed. These modelling assumptions influence the estimated magnitude of the delay reduction and therefore delimit the interpretation of the results.

The same fast-track model was also applied to the Mozilla data. Those estimates are not included in Table~\ref{tab:apache-fast-track}, however, because Mozilla's 31-day observation window is strongly right-censored. They are therefore treated only as directional evidence rather than as a comparable sensitivity analysis.

\textbf{Owner-constrained feasibility.} 
The Apache fast-track analysis above assumes that remediation capacity is fully fungible, such that any available worker can process any critical item. In practice, security findings are typically assigned to specific system owners, and remediation may be constrained by code ownership, authorisation, review, and release responsibilities. A separate analysis therefore relaxes this assumption by evaluating remediation capacity at the owner level. This owner-constrained analysis is performed for both Apache and Mozilla.
Under a 25\% reservation cap, sufficient owner-level capacity is available for 49.5\% of critical arrivals in Apache and 61.3\% in Mozilla. Increasing the cap to 50\% raises these shares to 82.5\% and 93.7\%, respectively. 
Increasing the reservation cap therefore substantially increases the share of critical work that can be accommodated within existing owner capacity. However, the Apache results also show that capacity reservation alone cannot resolve every bottleneck: 4.7\% of critical arrivals remain infeasible even when the responsible owner's full measured capacity is available because critical work arrives faster than that owner can process it. The effectiveness of capacity reservation therefore depends not only on the total amount of capacity available, but also on how that capacity is distributed across owners.
These findings indicate that aggregate remediation capacity alone is insufficient to characterise operational feasibility. Capacity may remain available elsewhere in the system while individual owners responsible for critical work are already saturated. The Apache fast-track estimates therefore represent an optimistic allocation scenario in which capacity can be deployed wherever demand arises, whereas the owner-level analysis provides a more realistic approximation of remediation feasibility under ownership constraints.

\textbf{Expertise-eligible borrowing.} 
The owner-level analysis shows that remediation capacity may remain available elsewhere in the system even when the responsible owner is saturated. Whether this capacity can be used depends, however, on whether other owners have demonstrated relevant expertise. Resolver overlap is therefore used as a proxy for cross-owner expertise eligibility, while recognising that previous involvement does not guarantee current availability or authorisation.
Allowing capacity to be borrowed from expertise-eligible owners produces markedly different effects in Apache and Mozilla. At a 25\% reservation cap, feasible critical arrivals increase from 49.5\% to 53.1\% in Apache, but from 61.3\% to 88.7\% in Mozilla. At a 50\% cap, feasibility increases from 82.5\% to 84.7\% in Apache and from 93.7\% to 99.3\% in Mozilla. These differences cannot be explained by the number of resolvers per owner alone, which is similar in the two systems (4.5 in Apache and 4.7 in Mozilla). Instead, the systems differ substantially in cross-owner connectivity. Apache owners have, on average, 0.4 eligible helpers and 75\% have no eligible helper, whereas Mozilla owners have 13.3 eligible helpers on average and only 6\% are isolated.
These findings indicate that the value of cross-owner capacity depends not only on whether spare capacity exists, but also on whether expertise links make that capacity accessible to the owners experiencing demand. The contribution of expertise connectivity cannot be quantified precisely in the present data because assignee information is available for only 73.5\% of Apache issues and 52.9\% of Mozilla issues, and the missingness may be systematic. More complete assignee histories are therefore required to estimate this effect robustly. Nevertheless, the observed contrast between Apache and Mozilla suggests that modelling cross-owner expertise as a connectivity network may help identify where otherwise inaccessible remediation capacity can be mobilised across ownership boundaries.

\textbf{Four observed mobility regimes.} The Apache--Mozilla comparison suggests that cross-owner borrowing is useful only when two conditions are met: owners must be connected through relevant expertise, and eligible helpers must have sufficient spare capacity. To examine whether these conditions generalise beyond the two primary datasets, the same owner-level analysis was applied to five public Jira organisations using projects as a common proxy for ownership \cite{ref18}. Table~\ref{tab:capacity-mobility} reports the resulting feasibility at a 25\% reservation cap.
The five organisations exhibit four distinct capacity-mobility regimes. Apache and Red Hat Jira show similarly limited cross-owner connectivity, with 75\% and 71\% of owners isolated, respectively. Under owner-only allocation, 49.5\% of critical arrivals are feasible in Apache and 51.4\% in Red Hat Jira. Expertise-eligible borrowing increases these shares only modestly, to 53.1\% and 55.2\%, corresponding to gains of 3.6 and 3.8 percentage points.

Qt shows a markedly different pattern. Although 25\% of owners have no eligible helpers, only 12.8\% of critical arrivals can be accommodated when each owner relies on its own available capacity. Allowing eligible owners to share spare capacity increases this share to 98.1\%, a gain of 85.3 percentage points. This suggests that Qt has substantial spare capacity, but much of it is located outside the owners receiving the critical work.

MongoDB shows a different case. All critical arrivals can already be accommodated by their responsible owners, so borrowing provides no additional benefit despite 6\% of owners having no eligible helpers. Cross-owner support is therefore unnecessary when sufficient capacity is already available where the critical work arrives.

MariaDB shows the opposite constraint. Every owner has at least one eligible helper, yet only 14.2\% of critical arrivals can be accommodated, and borrowing provides no improvement. Here, connections between owners exist, but the connected helpers do not have sufficient spare capacity to take on additional critical work.

\begin{table*}[!t]
\centering
\caption{Capacity mobility across five Jira organisations at a 25\% reservation cap. Each organisation contributes 2,000 issues; observation windows are non-identical.}
\label{tab:capacity-mobility}
\begin{tabularx}{\textwidth}{@{}lccccX@{}}
\toprule
Organisation
& \makecell{Owners without\\eligible helpers}
& \makecell{Owner-only\\feasibility}
& \makecell{Feasibility with\\borrowing}
& \makecell{Borrowing\\lift}
& \makecell[l]{Observed\\regime} \\
\midrule

Apache
& 75\%
& 49.5\%
& 53.1\%
& 3.6 pts
& Siloed \\

Red Hat Jira
& 71\%
& 51.4\%
& 55.2\%
& 3.8 pts
& Siloed \\

Qt
& 25\%
& 12.8\%
& 98.1\%
& 85.3 pts
& Connected with slack \\

MongoDB
& 6\%
& 100.0\%
& 100.0\%
& 0.0 pts
& Already provisioned \\

MariaDB
& 0\%
& 14.2\%
& 14.2\%
& 0.0 pts
& Connected but saturated \\

\bottomrule
\end{tabularx}
\end{table*}

Taken together, these cases show that remediation feasibility depends not only on how much capacity is available, but also on where that capacity is located and whether it can be transferred to the owners receiving critical work. Cross-owner borrowing is most valuable when local capacity is insufficient but spare capacity exists among owners with relevant expertise, as observed for Qt. It provides little additional value when sufficient capacity is already available locally, as in MongoDB, or when connected owners themselves lack spare capacity, as in MariaDB. Expertise connectivity and available capacity must therefore be considered jointly when assessing whether remediation work can be redistributed across an organisation.

The four observed regimes are descriptive rather than fixed characteristics of particular organisations. They represent different combinations of local capacity, cross-owner expertise connectivity, and available helper capacity. Mozilla, although analysed at component rather than project granularity, exhibits the same connected with slack pattern observed for Qt. Where critical work remains infeasible even after expertise-eligible borrowing, redistribution of existing capacity is insufficient; remediation would instead require additional capacity, reduced incoming demand, or expansion of the expertise available across owners.

\subsection{Network-Structured Exposure}\label{network-structured-exposure}

The preceding analyses demonstrate that organisational connectivity constrains the allocation of remediation capacity. To determine whether vulnerability exposure is similarly associated with technical connectivity, the analysis is extended to the npm package dependency graph.

Vulnerabilities show a weak but statistically significant tendency to occur in connected packages ($+0.083$, permutation $p = 0.011$). This association remains after accounting for the fact that highly connected packages naturally have more opportunities to be linked to other vulnerable packages. A temporal pattern is also observed. When vulnerabilities are documented in both a base package and one of its dependents, the vulnerability in the base package is recorded first in 60\% of cases ($p = 0.019$).

These findings have two implications. First, vulnerability exposure is not fully captured by considering packages independently. Dependency information may therefore complement item-level scoring by identifying connected parts of the software graph that warrant additional attention. Second, clustering along dependency relationships raises a potential limitation of the Poisson arrival assumption used in the fast-track queue model. Dependency-related vulnerabilities could produce correlated or bursty remediation demand rather than independent arrivals. The npm analysis does not measure remediation arrivals directly and therefore cannot establish such dependence, but it indicates that the Poisson assumption should be treated as a modelling approximation rather than an observed property of the remediation process.

The analysis is limited to a single software ecosystem and records documented vulnerabilities rather than exploit or remediation cascades. The observed temporal ordering therefore does not establish causal propagation from base packages to their dependents.

\section{Limitations}\label{limitations}
The analyses combine observational data, proxy measures, and model-based counterfactuals across several software ecosystems. Their interpretation therefore depends on the validity of the underlying measurements, modelling assumptions, and the extent to which the sampled systems represent broader remediation settings. The main limitations concern measurement, causal and modelling
assumptions, generalisability, and statistical interpretation.

\textbf{Measurement limitations.} Resolution time includes both time spent waiting for remediation capacity and time spent actively resolving an issue; these components cannot be separated directly in the public trackers. Resolution durations observed while queues are draining are therefore used only as a proxy for service time. Jira projects and Mozilla components are treated as proxies for ownership, while final assignees are used as proxies for resolvers. Resolver overlap captures evidence of prior experience across ownership boundaries, but does not establish current availability, authorisation, communication efficiency, or access to required review and release processes. Priority schemes also differ across communities and require organisation-specific normalisation. In the Red Hat dataset, dwell time measures the interval from public disclosure to the first Red Hat Security Advisory (RHSA), rather than issue-triage or implementation time. Finally, the npm analysis measures documented vulnerability presence and advisory timing; it does not observe exploit propagation or remediation cascades.

\textbf{Causal and modelling limitations.} All empirical analyses are observational and none constitutes a randomised intervention. The draining-transition analysis is based on only two treated Apache project queues, limiting the strength and generalisability of the difference-in-differences estimate. The fixed-capacity sequencing replay assumes that remediation effort is exchangeable across the reordered items. The fast-track analysis is an analytical counterfactual rather than a deployed intervention and depends on several queueing assumptions, including a pooled M/G/1 representation, Poisson arrivals, an estimated service-time distribution, and non-preemptive processing. The npm results further suggest that vulnerability exposure can be structured by dependency relationships, raising the possibility that remediation demand may not always arrive independently. This does not demonstrate a violation of the Poisson assumption in the Apache queue, but reinforces its interpretation as a modelling approximation.

The owner-level analyses introduce additional uncertainty. Missing assignee information may be systematic rather than random, and resolver overlap provides only a proxy for transferable expertise. The greedy allocation of helper capacity represents one feasible routing strategy and does not establish that the resulting allocation is optimal.

\textbf{Generalisability.} The primary evidence is drawn from two open-source issue trackers, one vendor security-advisory pipeline, and one package ecosystem. The capacity-mobility comparison extends the analysis to five public Jira organisations, each represented by 2,000 issues at project-owner granularity. Apache uses the longer-span primary sample, whereas MariaDB, Qt, MongoDB, and Red Hat Jira contain issues created from 2022 onwards. Differences in observation windows, priority schemes, ownership structures, and development processes limit direct comparison across organisations.

Public issue trackers also provide only partial representations of enterprise remediation processes. They do not capture private staffing data, informal ownership arrangements, authorisation boundaries, or internal escalation mechanisms. The four observed mobility regimes are therefore descriptive rather than universal organisational categories and require replication across additional organisations, time periods, ownership granularities, and the broader Jira corpus reported by Montgomery et al.~\cite{ref18}. The npm dependency graph likewise represents a single software ecosystem and should not be assumed to reproduce the technical coupling of enterprise systems.

\textbf{Statistical limitations.} Resolution times are strongly right-skewed, making estimates that depend on higher moments sensitive to a small number of very long-running items. Required service-time moments are therefore winsorised at the 99th percentile and sensitivity analyses are reported for the fast-track model. The governance comparison should also be interpreted cautiously because confidence intervals for adjacent systems overlap. Multiple analyses reuse the same underlying corpora and therefore do not constitute independent replications.

The network effects are statistically detectable but modest in magnitude and are estimated from a single dependency ecosystem. Similarly, the four capacity-mobility regimes summarise five observed organisations rather than estimate the prevalence of these regimes in a wider population. The reported numerical results should therefore be interpreted as evidence for the mechanisms examined in the sampled systems rather than as population-level estimates.

\section{Implications and Future
Work}\label{implications-and-future-work}
The results suggest that vulnerability remediation should be treated as a problem of workload, capacity, and allocation in addition to prioritisation. Across the analyses, remediation performance depends on whether sufficient capacity exists, where that capacity is located, and whether ownership and expertise constraints allow it to reach critical work. These findings have implications for both the operation of remediation systems and the design of future research.

\textbf{Implications for practice.}
Queue utilisation should be assessed before introducing more complex prioritisation or prediction models. When work arrives faster than it can be completed, improved ranking can change which items are processed first but cannot remove the underlying capacity constraint. The observed transitions from supercritical to draining Apache queues are consistent with substantially shorter remediation times once incoming work no longer exceeds completion capacity. Work-in-progress limits, demand control, or additional capacity may therefore be more consequential than further improvements in ranking accuracy when queues are persistently overloaded.

Other flow-control interventions may address persistent overload more directly. Work-in-progress (WIP) admission control can limit the amount of work entering an already congested queue, consistent with the observed reduction in delay following transitions from supercritical to draining queues and with prior evidence on Kanban-based WIP limits
\cite{ref11}. Staffing interventions could likewise increase capacity where persistent bottlenecks remain, although evaluating such policies requires workforce information that is not observable in the public datasets.

Capacity reservation provides a second operational lever, but its effectiveness depends strongly on the amount of capacity allocated. In the Apache fast-track model, small changes in reservation produce large differences in estimated critical-item delay as utilisation approaches saturation. Selecting a reservation level is therefore not simply a question of assigning more capacity to critical work: additional reservation reduces the capacity available for non-critical items. Remediation capacity allocation is consequently an optimisation problem in which critical delay must be balanced against congestion elsewhere in the system.

The location of capacity is equally important. Aggregate capacity can appear sufficient while individual owners remain unable to accommodate the critical work assigned to them. Cross-owner routing can address this imbalance only when two conditions are satisfied: relevant expertise must be reachable and eligible helpers must have spare capacity. The five-organisation comparison illustrates the consequences of these conditions. Borrowing is highly effective when both are present, as observed for Qt; unnecessary when sufficient capacity already exists with the responsible owners, as in MongoDB; and ineffective when expertise connections exist but helpers themselves lack capacity, as in MariaDB. Operational routing decisions should therefore consider capacity and expertise jointly rather than treating either as sufficient in isolation.

These findings also have implications for the use of severity scores. The relationship between severity and remediation speed varies substantially across the observed systems and is strongest in the more formally governed Red Hat errata process. Severity should therefore not be assumed to translate uniformly into faster remediation across organisational settings. Organisations should verify whether their prioritisation signals are actually reflected in remediation behaviour before relying on increasingly sophisticated scoring models.

Prediction systems require similar operational validation. Statistical coverage alone does not guarantee useful remediation deadlines: prediction intervals can widen sufficiently to preserve coverage while becoming too broad to support decision-making. Coverage and interval width should therefore be reported together. Abrupt widening may additionally provide an operational signal that the relationship between observed predictors and remediation times has changed, although the cause of such changes requires separate investigation.

Finally, remediation decisions may benefit from information beyond individual vulnerabilities. The npm analysis shows that documented vulnerabilities are weakly clustered along dependency relationships and tend to be recorded earlier in base packages than in their dependents. Dependency information may therefore complement item-level scoring by identifying connected parts of the software graph that warrant additional scrutiny following an upstream finding.

\textbf{Implications for research.}
The results motivate a shift from treating vulnerability remediation primarily as a ranking problem towards modelling the dynamics of the remediation system itself. Future work should examine the joint optimisation of sequencing, capacity reservation, and routing under ownership and expertise constraints. The present analyses evaluate these mechanisms separately; a more complete model could determine how scarce remediation capacity should be distributed across competing classes of work and eligible owners while controlling delays elsewhere in the system.

More dynamic allocation policies also warrant investigation. For example, Whittle-index policies \cite{ref12} could prioritise individual remediation tasks as their states evolve, but require reliable estimates
of item-level state transitions. The available issue histories do not provide sufficient information to estimate such transition models.

Stronger causal evidence is also required. The draining-transition analysis contains only two treated Apache queues, while the sequencing and fast-track analyses are counterfactual or model-based. Controlled field studies could test whether reserving capacity for critical work reduces long-duration delays without producing unacceptable spillover to non-critical work. Similar experiments could evaluate whether expertise-aware routing improves remediation feasibility and resolution time in operational settings.

More complete workforce and assignment data would permit a more direct characterisation of capacity mobility. Bugzilla assignment and component histories could support time-varying analyses of routing, ownership changes, and coordination delays, with explicit treatment of temporal biases such as immortal-time bias. Data containing team size, workload, authorisation boundaries, and skill histories would further allow resolver connectivity and helper availability to be measured directly rather than inferred from historical assignee overlap.

The four observed capacity-mobility regimes should be tested beyond the five organisations considered here. The analysis can be extended to suitable instances in the 16-Jira corpus \cite{ref18}, repeated across multiple observation windows, and evaluated using consistent ownership definitions and organisational covariates. Such replication would establish whether the observed combinations of local capacity, expertise connectivity, and helper slack recur systematically across organisations.

The governance-associated severity gradient likewise warrants broader investigation. Replication across organisations with different remediation processes could test whether stronger links between severity and remediation speed systematically accompany more formal prioritisation and accountability mechanisms, or instead arise from other organisational differences.

Finally, the network analysis motivates two extensions. First, dependency-aware remediation models could test whether upstream vulnerability information improves identification of downstream exposure beyond item-level risk scores. Second, vulnerability clustering along dependency relationships raises the possibility that remediation demand may be temporally correlated rather than independently arriving. Future queueing analyses should therefore test arrival dependence directly and compare the present Poisson approximation with models that allow bursty or correlated demand.

\section{Reproducibility}\label{reproducibility}

All analyses are based on public datasets and fixed local caches to support deterministic reruns. The empirical results, counterfactual analyses, and queueing calculations are implemented in a reproducible analysis pipeline. Key numerical results are independently recomputed where possible, including the Hill tail estimate, severity-to-speed discrimination, conformal coverage,
capacity-reservation sensitivity, and capacity-mobility measures.

\section{Conclusion}\label{conclusion}

This study shows that vulnerability remediation cannot be understood solely as a prioritisation problem. Across the systems examined, remediation performance is closely tied to queue state, available capacity, ownership constraints, and the ability to move work to resolvers with relevant expertise. More sophisticated ranking or prediction cannot remove these constraints when work arrives faster than it can be processed or when capacity is unavailable where critical work occurs.

The results therefore support a broader view of vulnerability management as a flow-control problem. Sequencing, capacity reservation, and cross-owner routing provide distinct mechanisms for improving the movement of critical work, but their effectiveness depends on the operational structure in which they are applied. As vulnerability discovery accelerates, effective remediation will increasingly require not only identifying what matters most, but ensuring that sufficient and appropriately located capacity exists to act on it.

\end{document}